\documentclass[11pt,twocolumn,aps,prl,superscriptaddress,reprint]{revtex4-1}

\usepackage[abs]{overpic}
\usepackage{graphicx,graphics}
\usepackage{dcolumn}
\usepackage{latexsym,verbatim}
\usepackage{bm}
\usepackage{color}
\usepackage[dvipsnames]{xcolor}
\usepackage{ulem}
\usepackage[framemethod=default]{mdframed}
\usepackage[breaklinks=true,colorlinks,citecolor=blue,linkcolor=blue,urlcolor=blue]{hyperref}
\usepackage[makeroom]{cancel}
\usepackage{fancybox}
\usepackage{siunitx}

\newcommand{\addDB}[1]{\textcolor{black}{#1}}

\begin{document}
{
	\makeatletter
	\def\frontmatter@thefootnote{%
		\altaffilletter@sw{\@fnsymbol}{\@fnsymbol}{\csname c@\@mpfn\endcsname}%
	}%
	\makeatother

\title[THz]{Observation of terahertz-induced magnetooscillations in graphene}		

\author{Erwin M\"onch}
\affiliation{Terahertz Center, University of Regensburg, 93040 Regensburg, Germany}


\author{Denis A. Bandurin$^*$}
\affiliation{Department of Physics, Massachusetts Institute of Technology, Cambridge, Massachusetts 02139, USA.}

\author{Ivan A. Dmitriev}
\affiliation{Terahertz Center, University of Regensburg, 93040 Regensburg, Germany}
\affiliation{Ioffe Institute, 194021 St. Petersburg, Russia}

\author{Isabelle Y. Phinney}
\affiliation{Department of Physics, Massachusetts Institute of Technology, Cambridge, Massachusetts 02139, USA.}

\author{Ivan Yahniuk}
\affiliation{CENTERA, Institute of High Pressure Physics PAS, 01142 Warsaw,  Poland}

\author{Takashi Taniguchi}
\affiliation{Advanced Material Laboratory, National Institute of Material Science, Tsukuba, Ibaraki 305-0044, Japan}

\author{Kenji Watanabe}
\affiliation{Advanced Material Laboratory, National Institute of Material Science, Tsukuba, Ibaraki 305-0044, Japan}

\author{Pablo Jarillo-Herrero}
\affiliation{Department of Physics, Massachusetts Institute of Technology, Cambridge, Massachusetts 02139, USA.}

\author{Sergey D. Ganichev$^*$}
\affiliation{Terahertz Center, University of Regensburg, 93040 Regensburg, Germany}

\begin{abstract}
When high-frequency radiation is incident upon graphene subjected to a perpendicular magnetic field, graphene absorbs incident photons by allowing transitions between nearest LLs that follow strict selection rules dictated by angular momentum conservation. Here we show a qualitative deviation from this behavior in high-quality graphene devices exposed to terahertz (THz) radiation. We demonstrate the emergence of a pronounced THz-driven photoresponse, which exhibits low-field magnetooscillations governed by the ratio of the frequency of the incoming radiation and the quasiclassical cyclotron frequency. We analyze the modifications of generated photovoltage with the radiation frequency and carrier density and demonstrate that the observed photoresponse shares a common origin with microwave-induced resistance oscillations previously observed in GaAs-based heterostructures, yet in graphene, it appears at much higher frequencies and persists above liquid nitrogen temperatures. Our observations expand the family of radiation-driven phenomena in graphene and offer potential for the development of novel optoelectronic devices.

\begin{center}
\textbf{*Email:} bandurin@mit.edu; sergey.ganichev@ur.de 
\end{center}
\end{abstract}

\maketitle	

In recent years, graphene has provided access to a rich variety of quantum effects, owing to its nontrivial band topology, quasi-relativistic energy spectrum, and 
high electron mobility~\cite{CastroNetoRMP}. These properties also determine the unique response of graphene to external electromagnetic fields: a 
host of interesting magneto-optical phenomena such as giant Faraday rotation~\cite{Kuzmenko2010}, radiation-driven nonlinear transport~\cite{Glazov2014}, gate-tunable magnetoplasmons~\cite{Magnetoplasmons1,Magnetoplasmons2}, ratchet \cite{Olbrich2016} and magnetic quantum ratchet effects~\cite{ratchet},  as well as colossal magneto-absorption~\cite{LLspectroscopyKim2007,Kuzmenko2019}, to name a few, has
been  discovered in graphene exposed to infrared and THz radiation. At these radiation frequencies, graphene also supports the propagation of long-lived gate-tunable plasmon-polaritons~\cite{Juplasmon,THsplamonsSNOM,GrPlasmonicsMarco,Bandurin_THz_NComm} enabling ultra-high confinement of electromagnetic energy and offering a platform for the fundamental studies of radiation-matter interactions at the
nanoscale~\cite{NonlocalTHZSNOM,GNI2018}. Furthermore, in addition to fundamental interest, graphene's conical band structure together with fast carrier thermalization has generated much excitement for practical photonic and optoelectronic applications~\cite{Ferrari_review}, particularly for ultra-long wavelengths~\cite{THzreviewAPL,koppens_detectors,Microwave2,KC_Microwave}. Graphene-based devices have been shown to perform as highly-sensitive, ultra-fast, and broadband detectors of THz radiation, a technologically challenging frequency domain to which other materials struggle to respond~\cite{THzreviewAPL,koppens_detectors}. 

In this paper, we uncover a different kind of radiation-driven phenomena in graphene by studying the interplay of THz absorption with electron transport. We demonstrate that graphene subjected to a perpendicular magnetic field, $B$, exhibits pronounced $1/B$ magnetooscillations in response to incident THz-radiation. These  magnetooscillations emerge in the range of weak magnetic fields well below those needed for the full development of conventional Shubnikov-de Haas oscillations, and their fundamental frequency is found to be proportional to the frequency of incoming radiation, $f$. By analyzing the observed THz-driven photovoltage as a function of gate-induced carrier density, $n$, we show that the resonant condition appears when $2\pi f$ is commensurate with the frequency of the electron's quasiclassical cyclotron motion, $\omega_c$. The latter indicates that the observed photovoltage oscillations have the same physical origin as the
microwave-induced resistance oscillations (MIRO) in high-mobility 2DES with parabolic spectrum \cite{Zudov2001,ye:2001,mani:2002,zudov:2003,yang:2003,RevModPhysVanya}. However, in graphene, they emerge at higher frequencies and, quite remarkably, persist above liquid nitrogen temperatures, $T$, while their frequency is tunable by the gate voltage. Our findings extend the family of radiation-driven effects in graphene, paving the way for future studies of nonequilibrium electron transport and revealing new avenues for the development of 
terahertz optoelectronic devices. 

\begin{figure*}[ht!]
	\centering
	\includegraphics[width=0.8\textwidth]{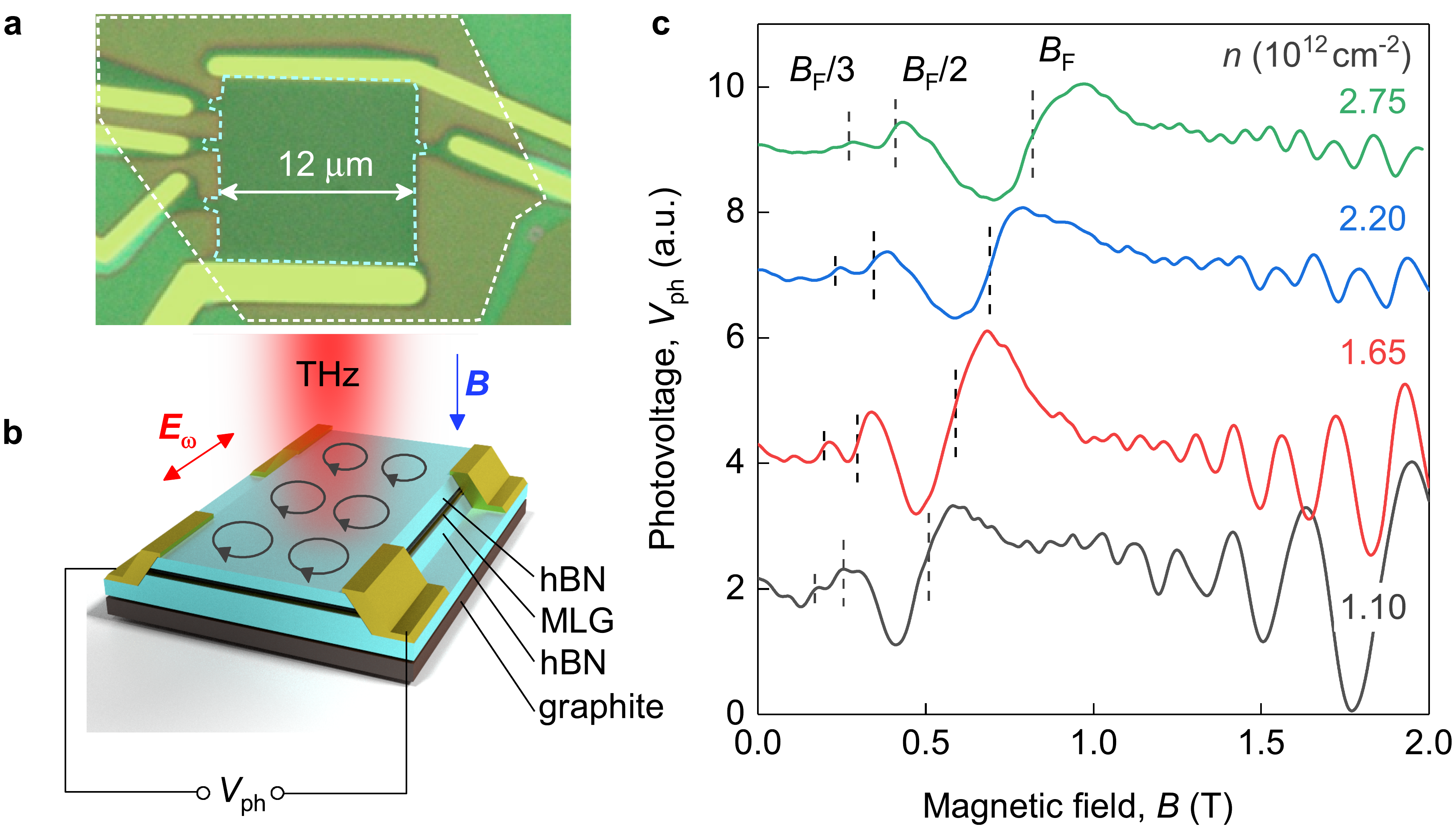}
	\caption{ (a) Optical photograph of one of our encapsulated devices. Dashed blue line highlights the shape of graphene Hall bar. Dashed white line contours the graphite bottom gate region. (b) Measurement configuration:  a normally incident THz laser radiation is focused on the graphene  device placed in an out-of-plane magnetic field, $B$. Circles illustrate the electron cyclotron motion. Red and blue arrows show the in-plane polarization of the incident radiation and out-of-plane magnetic field.  (c) Examples of the photovoltage dependence on magnetic field recorded in response to 0.69 THz laser radiation at $T=4.2$~K. Different traces, corresponding to given carrier densities, $n$, are up shifted for clarity. Dashed vertical lines at $B_F/N$, $N=1,2,3$, indicate harmonics of the cyclotron resonance in graphene, see Eq.~(\ref{BF}). 
 } \label{Fig1}
\end{figure*}

\textbf{Devices and measurements}. Our samples are multi-terminal devices made of monolayer graphene encapsulated between two relatively thick ($\sim50$~nm) crystals of hexagonal boron nitride (hBN) fabricated using a high-temperature release method~\cite{Hot-transfer_NComm}. The devices were 
\addDB{patterned in a conventional Hall bar geometry} and transferred on top of a 20 nm thick graphite flake, Fig.~\ref{Fig1}a-b  (See Supporting Information for details). The graphite served as a gate electrode, by which $n$ was controlled, and was also used to screen remote charged impurities in the Si/SiO$_2$ substrate~\cite{Sahsa_Nature}. The devices \addDB{had  a width} of \SI{12}{\micro\meter} 
and exhibited high mobility, $\mu$, 
exceeding $3\times10^5$~cm$^2/$Vs 
at liquid helium $T$. 



Experiments were performed in a variable temperature optical cryostat equipped with a polyethylene window to allow coupling of the sample with linearly-polarized THz radiation. The latter was generated by a continuous wave molecular gas laser operating at frequencies $f=0.69$ and 1.63 THz with radiation power up to 20 mW \cite{Danilov2009,Olbrich2013}. By using a pyroelectric camera \cite{Ziemann2000}, the laser spot with diameter about 2.5 mm was guided to the center of the device. The THz beam was modulated by an optical chopper operating at a frequency of about 80 Hz. Photoresponse measurements were carried out using a standard lock-in technique: the photovoltage, $V_\mathrm{ph}$, was recorded as the phase-locked potential difference generated in response to the chopper-modulated THz radiation, between a pair of contacts. All data were obtained in the Faraday configuration (Fig.~\ref{Fig1}b) with both the laser beam and magnetic field oriented perpendicular to the graphene plane.




\begin{figure*}[!ht]
	\centering
	\includegraphics[width=1\textwidth]{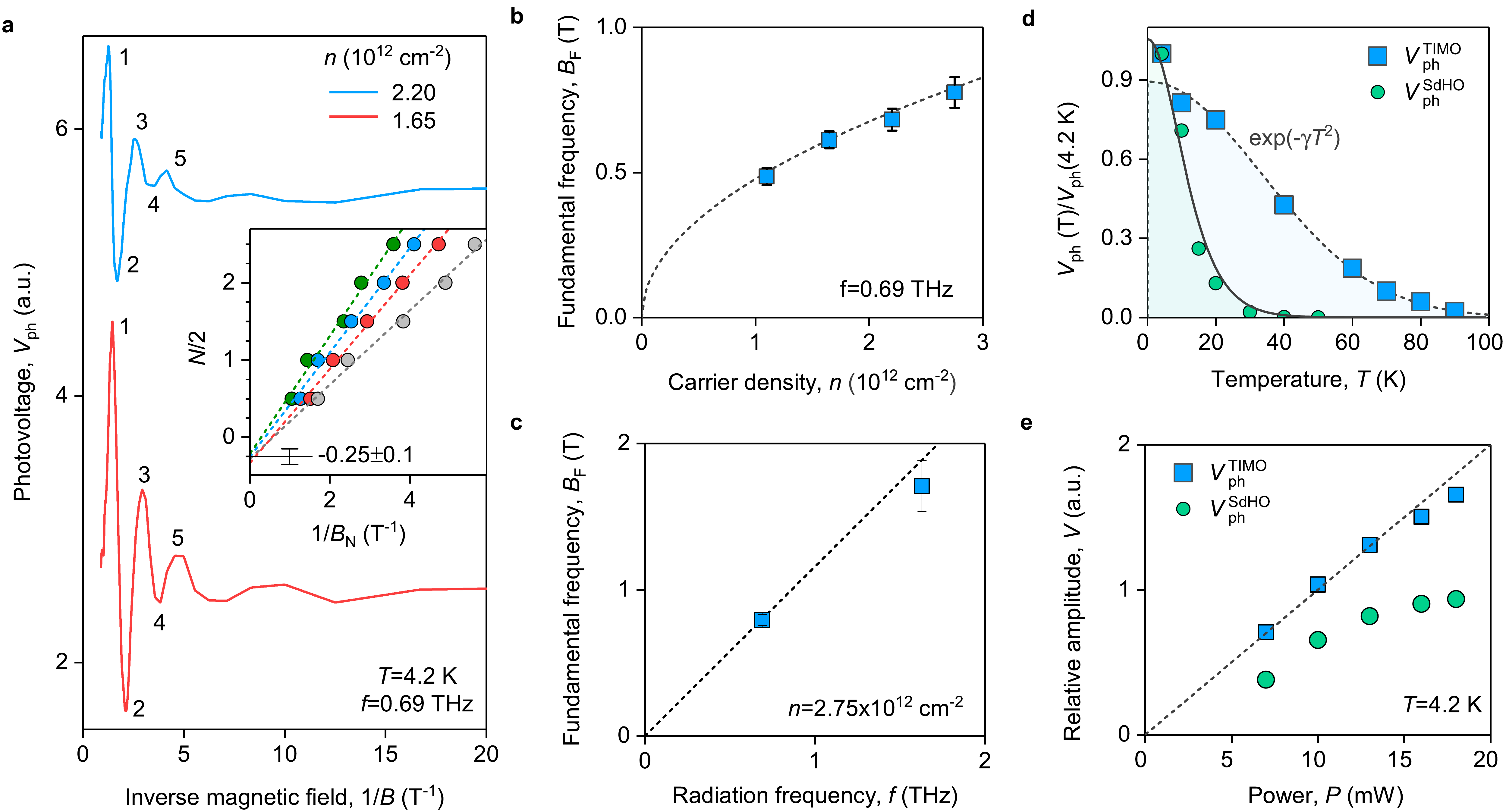}
	\caption{(a) Examples of TIMO in $V_\mathrm{ph}$ as a function of inverse magnetic field for given $n$. Inset: TIMO extrema indices $N=1,2,...5$ (as marked) against the inverted values $1/B_N$ of magnetic field at which they appear. Color-coding is the same as in Fig. 1c. Dashed lines: linear fits of the data used to determine the fundamental frequency $B_\mathrm{F}$ (the slope) and the phase of TIMO, see Eq.~(\ref{phase}). (b) $B_\mathrm{F}$ as a function of $n$ for given $f$. (c) $B_\mathrm{F}$ versus $f$ for given $n$. Dashed lines in (b-c): plots of $B_\mathrm{F}(n, f)$ according to Eq.~(\ref{BF}) with $v_\mathrm{F}=(1.07\pm0.12)\times 10^6$ m/s.
	(d) Temperature dependencies of the amplitude of the first period of TIMO ($V_\mathrm{ph}^\mathrm{TIMO}$, blue symbols) and of the SdHO-periodic photovoltage ($V_\mathrm{ph}^\mathrm{SdHO}$, green symbols) 
	for $n=2.75\times 10^{12}$ cm$^{-2}$. Solid black line: Fit of $V_\mathrm{ph}^\mathrm{SdHO}$ to the Lifshitz-Kosevich formula  yielding $m_\mathrm{c}\approx0.03m_\mathrm{e}$ where $m_\mathrm{e}$ is the free electron mass. Dashed black line: fit of $V_\mathrm{ph}^\mathrm{TIMO}$ 
	for $T\geq 10$~K with the function $\exp(-\gamma T^{2})$ yielding $\gamma=4.4\times 10^{-4}$\,K$^{-2}$. (e) Incident THz power dependencies of $V_\mathrm{ph}^\mathrm{TIMO}$ (blue symbols) and $V_\mathrm{ph}^\mathrm{SdHO}$ (green symbols). 
	Dashed black line: linear fit to 
	the data.
	}
	\label{Fig2}
\end{figure*}

The central result of our study is presented in Fig.~\ref{Fig1}c, which shows the emergence of $V_\mathrm{ph}$ in response to $f=0.69$ THz radiation when a magnetic field, $B$, perpendicular to graphene  is applied. Different traces correspond to several representative values of $n$. Two kinds of magnetooscillations are clearly distinct in the data. At $B>$\,\SI{1}{\tesla}, $V_\mathrm{ph}$ exhibits fast $1/B$-periodic oscillations which display periodicity of conventional Shubnikov-de Haas oscillations (SdHO). The presence of SdHO-periodic signals in the photovoltage is not surprising and is in line with previous observations \cite{Nernst_graphene_SDHO, HgTe_SdHO_response}.  Strikingly, at lower $B<$\,\SI{1}{\tesla}, where the SdHO-periodic oscillations get exponentially suppressed, another distinct magneto-oscillation pattern emerges. These low-$B$ oscillations, for brevity denoted THz-induced magnetooscillations (TIMO), will be explored in the remainder of this paper.

In Fig.~\ref{Fig2}a, we replot two examples of $V_\mathrm{ph}$ from Fig.~\ref{Fig2}a as a function of inverse magnetic field. Both traces clearly indicate the $1/B$-periodicity of TIMO. This periodicity is further verified by plotting the indices $N=$ 1, 2, \ldots 5 of the consecutive peaks and dips, see Fig.~\ref{Fig2}a, against the values $1/B_\mathrm{N}$ of the inverse magnetic field at which they appear. As demonstrated in the inset of Fig.~\ref{Fig2}a, for each $n$ in Fig.~\ref{Fig1}c, the positions of all extrema fall onto straight lines. The slope of these lines yields the fundamental frequency of TIMO, $B_\mathrm{F}$, which varies with $n$ and $f$, as shown in Figs.~\ref{Fig2}b-c (for further details see Supporting Information). Moreover, all lines cross the vertical axis at the same point $-0.25\pm 0.1$. This behavior yields the relation
 $N/2=B_\mathrm{F}/B_\mathrm{N}-1/4$, which translates into
 \begin{equation}
     V^\mathrm{TIMO}_\mathrm{ph}\propto -\sin(2\pi B_\mathrm{F}/B)\,,
     \label{phase}
 \end{equation}
 and thus establishes the TIMO phase, which will be important for the further analysis.
 
We have also studied $V_\mathrm{ph}(B)$ at different temperatures, $T$, and found that the amplitudes of SdHO-periodic oscillations and TIMO exhibit very different 
\addDB{$T$-dependences} as shown in Fig.~\ref{Fig2}d (for further details see Supporting Information). Remarkably, we observe that TIMO, and in particular their first period, can be well resolved even above liquid nitrogen temperatures, in sharp contrast to the SdHO-periodic signal, which vanishes completely at $T\sim40$~K over the entire $B-$ and $n-$ranges in which the measurements were performed. The latter dependence can be well fitted by the conventional Lifshitz-Kosevich formula \cite{shoenberg:1984,ando:1982a} as illustrated by the solid line.
 
The evolution of TIMO and SdHO-periodic signals with the power of incident THz radiation, $P$, is also found to be different, see Fig.~\ref{Fig2}e. The $P$-dependence of the TIMO amplitude is fairly linear, with a weak tendency to saturation at highest $P$\SI{>10}{\milli\watt}. This means that much stronger TIMO can be observed using more powerful THz sources. In contrast to TIMO, the amplitude of SdHO-periodic photoresponse clearly displays a sublinear $P$-dependence, which may reflect the electron heating caused by the THz irradiation.

\textbf{MIRO physics in graphene devices.} Below we argue that the above experimental results identify  TIMO as a graphene analogue of microwave-induced resistance oscillations (MIRO) \cite{Zudov2001,ye:2001,mani:2002,zudov:2003,yang:2003,smet:2005,RevModPhysVanya,zudov:2014,yamashiro:2015,karcher:2016,zadorozhko:2018,otteneder:2018,monarkha:2019}.
Experiments on high-mobility 2DES 
demonstrated that illumination with microwaves can lead to the emergence of strong magnetooscillations in static longitudinal resistance \cite{Zudov2001,ye:2001,mani:2002,zudov:2003,yang:2003,RevModPhysVanya}.
The maxima and minima of MIRO 
appear around the positions of harmonics of the cyclotron resonance (CR) given by  $\omega= M \omega_c$, where $\omega=2\pi f$, $M=1,\,2,\ldots$, and $\omega_c=e B/m$ is the cyclotron frequency \cite{mani:2004e}. It is thus natural to
attribute this effect to resonant photon-assisted transitions between distant Landau levels (LLs)\cite{abstreiter:1976,fedorych:2010,ando:1975a,dmitriev:2003,briskot:2013}.
Such processes require simultaneous impurity scattering 
\cite{ando:1975a,dmitriev:2003,briskot:2013}, since in the absence of disorder only 
transitions between neighboring LLs ($M=1$) are dipole-allowed.
Indeed, MIRO are observed in the range of $B$ where LLs are strongly broadened by disorder, see Fig.~\ref{Fig3}a. This is reflected in an exponential decay towards low $B$ that is similar to the SdHO and other quantum corrections to the classical Drude-Boltzmann transport coefficients \cite{RevModPhysVanya,ando:1982a}. 

In order to understand how the above photon-assisted transitions between broadened LLs lead to magnetooscillations in static transport observables such as photoconductivity \cite{dmitriev:2009b} and photovoltage \cite{dmitriev:2009a}, a theoretical framework involving two closely related mechanisms, often reffered to as displacement \cite{Ryzhii,Durst,VavilovAleiner,khodas:2008} and inelastic \cite{dmitriev:2003,dmitriev:2005}, has been developed. 
A hallmark of both mechanisms is that the effect vanishes at exact positions of the CR harmonics. We illustrate this in Fig.~\ref{Fig3}a, which represents the displacement mechanism. Solid lines show the maxima of the local density of states (DOS) in broadened LLs (shown on the left) which are tilted in the presence of a static electric field $\textbf{\textit{E}}$.  In a magnetic field, 
any impurity scattering is accompanied by a real-space displacement $\Delta \textit{\textbf{X}}$ of the center of the electron cyclotron orbit. In the example of Fig.~\ref{Fig3}a, the photon energy $\hbar\omega$ slightly exceeds the energy separation $\Delta \varepsilon$ between the involved LLs. This defines the preferred direction of the displacement $\Delta \textbf{\textit{X}}$ due to the photon-assisted impurity scattering, and, consequently, the oppositely directed contribution to the photocurrent $\textbf{\textit{j}}_\mathrm{\textbf{ph}}$. 
As one would expect from golden rule arguments \cite{Ryzhii,RevModPhysVanya}, the displacement vector points to the right, towards the maximum in the local DOS associated with the $K+2$ LL. The direction of $\Delta \textbf{\textit{X}}$ would reverse for the opposite sign of $\hbar\omega-\Delta \varepsilon$ and, therefore, the nonequilibrium current $\textbf{\textit{j}}_\mathrm{\textbf{ph}}$ can flow both along and against $\textbf{\textit{E}}$, depending on the sign of $\hbar\omega-\Delta \varepsilon$, and vanishes at the positions of CR harmonics.

\begin{figure*}[!ht]
	\centering
	\includegraphics[width=0.8\textwidth]{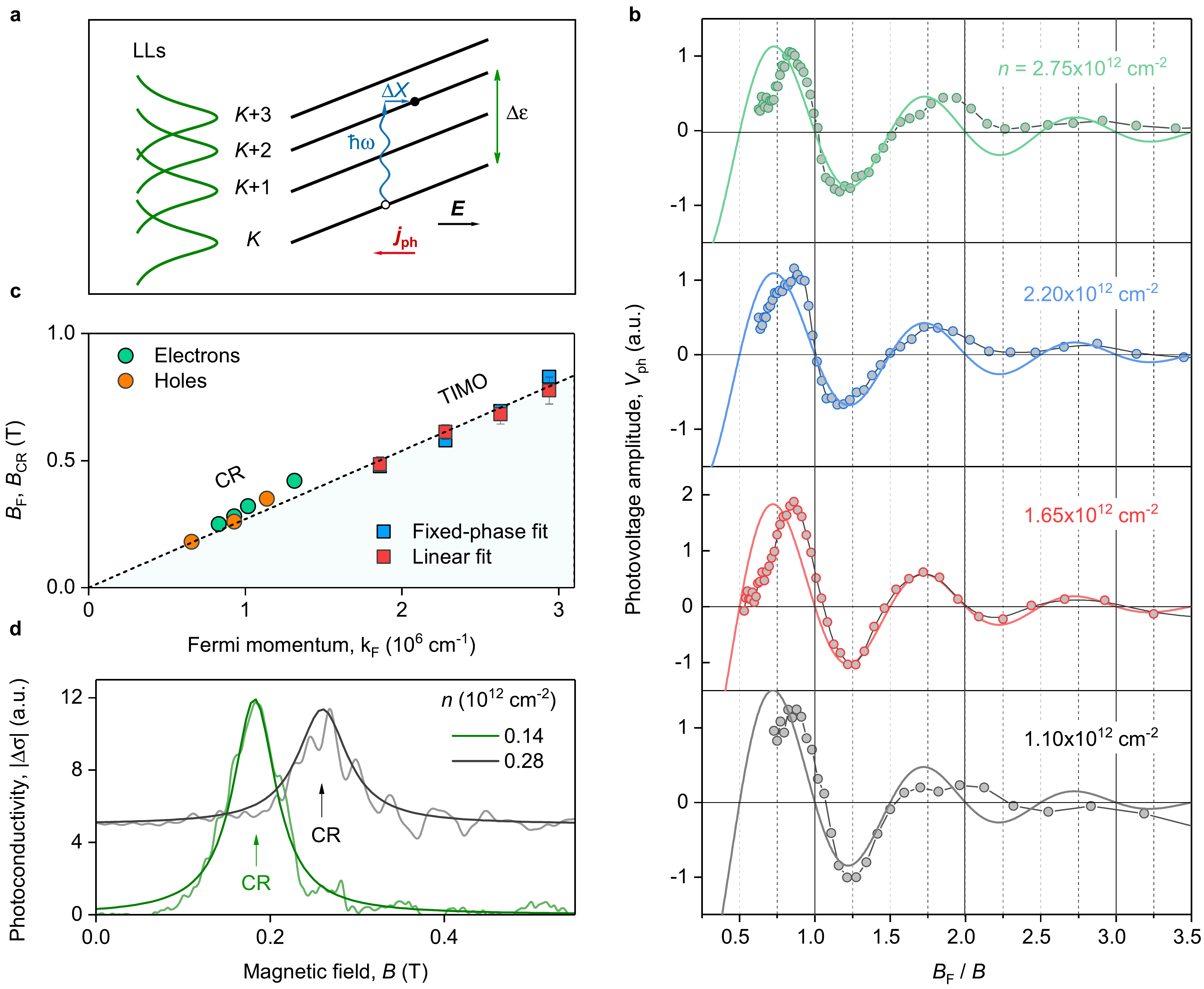}
	\caption{ (a) Schematic for the displacement mechanism of MIRO. Solid lines show how the energy positions of the centers of broadened LLs (shown on the left) are shifted in real space in the presence of a static electric field $E$. The impurity-assisted photon absorption leads to an inter-LL transition accompanied by a real space displacement of electron's cyclotron orbit, $\Delta \textbf{\textit{X}}$. The preferred direction of $\Delta \textbf{\textit{X}}$ is sensitive to the density of available for scattering states. On the schematic, the photon energy $\hbar\omega$ slightly exceeds the distance $\Delta\varepsilon$ between the involved LLs, and the density of final states is larger for the spatial shifts in the direction of $\textit{\textbf{E}}$. This results in generation of the photocurrent $\textbf{\textit{j}}_\mathrm{\textbf{ph}}$ in the opposite direction.  (b) Low-field $V_\mathrm{ph}$ as a function of $B_\mathrm{F}/B$ after subtraction of smooth non-oscillating background (circles). Solid lines: fits to Eq.~(\ref{Vph}).  The value of $B_\mathrm{F}$  for each carrier density is chosen to provide the best fit for the period of TIMO, with the phase fixed by Eq.~(\ref{Vph}). (c) $B_\mathrm{F}$ as a function of $k_\mathrm{F}$ determined by the fixed-phase fits to Eq.~\ref{Vph} (blue squares) and the linear fits in Fig.~\ref{Fig2}a (red squares). Dashed line shows the fit of $B_\mathrm{F}$ values obtained via both methods to $B_\mathrm{F}=\hbar\omega k_\mathrm{F}/e v_\mathrm{F}$, yielding $v_\mathrm{F}=(1.07\pm0.12)\times 10^6$ m/s. Circles mark the positions $B_\mathrm{CR}$ of the CR peaks detected in the photoconductivity at small $|n|$, see panel (d).  (d) Examples of photoconductivity $|\Delta \sigma|$ vs $B$ dependencies demonstrating pronounced quasi-classical CR  at given $n$. Solid lines: Lorentzian fits of the data used to determine $B_\mathrm{CR}$ (arrows) shown in panel (c). The curves are shifted for clarity. $T=4.2$ K.   }
	\label{Fig3}
\end{figure*}

In addition to $\textbf{\textit{j}}_\mathrm{\textbf{ph}}$ associated with such  displacements of orbits, the resonant inter-LL transitions lead to an unusual modification of the Fermi-Dirac energy distribution of electrons, which acquires a nonequilibrium correction proportional to the oscillatory DOS \cite{dmitriev:2003,dorozhkin:2016}. Since the amplitude of oscillations in the energy distribution is controlled by inelastic scattering processes, the corresponding contribution to MIRO \cite{dmitriev:2005} is termed inelastic. A combination of the displacement and inelastic mechanisms has successfully explained most experimental findings associated with MIRO~\cite{RevModPhysVanya}.

In 2DES with a parabolic energy dispersion,
the LL spectrum is equidistant, and MIRO display $2\pi$-periodicity with the ratio $\omega/\omega_c\equiv m\omega/e B$. The fundamental frequency of MIRO, $B_F=m \omega/e$, is thus insensitive to variations of the electron density, neglecting small changes of the effective mass $m$ due to density-dependent renormalization induced by electron-electron interactions~\cite{MIRO_MASS}.
This property should clearly change in graphene featuring a non-equidistant spectrum of LLs, $E_{K}\propto\sqrt{K}$, $K=0$, 1, 2$\ldots$~\cite{ManiTHzGraphene}. 
In a narrow energy window of $\mathrm{max}\{\hbar\omega, k_\mathrm{B} T\}\ll E_\mathrm{F}$ around the chemical potential $E_\mathrm{F}$, which is available for impurity-assisted emission and absorption of photons, the spacing between LLs at relevant 
$K\gg 1$ can be still approximated as
$\hbar\omega_c=\hbar e B/m_\mathrm{c}$, but with the density-dependent cyclotron mass $m_\mathrm{c}=\hbar\sqrt{\pi n}/v_\mathrm{F}$ \cite{Kostya2005}. Therefore, the fundamental frequency of TIMO,
\begin{equation}
    \label{BF}
    B_\mathrm{F}=\hbar\omega\sqrt{\pi n}/e v_\mathrm{F},
\end{equation}
is expected to scale with $\sqrt{n}$. These expectations are in excellent agreement with our findings: dashed lines in Fig.~\ref{Fig2}b,\,c demonstrate that the fundamental frequency of TIMO accurately follows Eq.~(\ref{BF}) as a function of $n$ and $\omega$ if one uses $v_\mathrm{F}=1.07\times10^6~\mathrm{m/s}$ typical for the relevant range of densities in graphene \cite{Kostya2005,Kim2005,kumaravadivel:2019}.

TIMO also displays other common features with MIRO, including the vanishing photoresponse at integer $B_\mathrm{F}/B$ (see vertical dashed lines in Fig.~\ref{Fig1}c), and an exponential damping towards low B. These further
manifestations of the MIRO phenomenon are evident from Fig.~\ref{Fig3}b, which compares the TIMO data 
with the conventional MIRO waveform,
\begin{equation}
V_\mathrm{ph}=- A \exp(-\alpha B_\mathrm{F}/B) \sin(2\pi B_\mathrm{F}/B),
\label{Vph}
\end{equation}
typical for weak oscillations in the regime of strongly overlapping LLs and small radiation intensity \cite{shi:2017,RevModPhysVanya}. The solid lines in Fig.~\ref{Fig3}b are fits to Eq.~(\ref{Vph}) with constants $A$, $B_\mathrm{CR}$, and $\alpha$ used as fitting parameters. This treatment complements the procedure presented in Fig.~\ref{Fig2}a, where the phase of TIMO was not fixed as in Eq.~(\ref{Vph}) but rather emerged as a result of the analysis, see Eq.~(\ref{phase}). 
The values of $B_\mathrm{F}$, obtained using either of the two fitting procedures, nearly coincide. They are plotted together as a function of the Fermi momentum $k_\mathrm{F}=\sqrt{\pi n}$ in Fig.~\ref{Fig3}c, and exhibit the proportionality to $k_\mathrm{F}$ in accordance with Eq.~(\ref{BF}). A fit for the slope yields the value $v_\mathrm{F}=(1.07\pm 0.12)\times 10^6$~m/s, in good agreement with previous studies~\cite{CastroNetoRMP}. 

Our observations establish that despite their resonant character, TIMO vanish at the exact position $B=B_\mathrm{F}$ of the cyclotron resonance (CR). This makes them markedly different from more conventional effects in the photoresponse, which are related to the resonant heating of electrons due to enhanced Drude absorption near the CR. 
Such CR-enhanced photoresponse was also detected in our devices, but at small $|n|$, where TIMO were not observed. Two examples of the photoconductivity, $\Delta \sigma$, traces featuring the CR-centered peaks are shown in Fig.~\ref{Fig3}d. The positions $B_\mathrm{CR}$ of these CR peaks are included in Fig.~\ref{Fig3}c. They fall onto the dashed line representing $B_\mathrm{F}(k_\mathrm{F})$ dependence extracted from TIMO, and further substantiate the previous analysis. 

\textbf{Damping of THz-induced magnetooscillations in graphene.} We now focus our attention on the low-$B$ damping and $T$-dependence of TIMO, which turn out to be closely
interrelated.
Fitting the low-$T$ TIMO data using Eq.(\ref{Vph}), see Fig.~\ref{Fig3}b, we find that the low-$B$ damping of TIMO is well reproduced by the factor $\exp(-\alpha B_\mathrm{F}/B)$ with $\alpha\simeq 1$. Within the displacement and inelastic theoretical frameworks~\cite{dmitriev:2009b}, this factor describes an increasing overlap of the broadened LLs upon lowering $B$, and can be rewritten as the square, $\delta^2$, of the conventional Dingle factor $\delta=\exp(-\pi/\omega_\mathrm{c}\tau_\mathrm{q})$. Remarkably, the corresponding value of the quantum scattering time, $\tau_\mathrm{q}=1/\alpha f\simeq 1.5$\,ps, 
significantly exceeds the values $\tau_\mathrm{q}\sim 0.3$\,ps extracted from the SdHO measurements in graphene samples of similar quality \cite{zeng:2019}, yet are a few times smaller than typical low-$T$ values $\tau_\mathrm{tr}\sim 5\div10$\,ps of the transport scattering times in our samples. The obtained scattering times are at least an order of magnitude shorter than those of GaAs-based heterostructures used to study MIRO~\cite{RevModPhysVanya}. In view of the relation $\alpha=1/f\tau_\mathrm{q}$, this necessitates the use of higher (THz) frequencies in graphene to observe non-equilibrium phenomena of this type. On the other hand, calculations show that an increase of the radiation frequency causes a very fast $f^{-4}$ decay in MIRO amplitude $A$, as opposed to a slower $f^{-2}$ decay of the Drude absorption and associated electron heating~\cite{RevModPhysVanya}. The observation of the MIRO-like oscillations in conventional 2DES thus proved challenging for $f$ above 1 THz \cite{TIRO,TIRO2}. Our samples, in contrast, revealed clear signatures of TIMO at elevated THz frequencies (see Fig.~\ref{Fig2}c and Supporting Information), which points to a exceptional stability of TIMO against heating effects.

Anomalously slow $T$-decay of TIMO is also not less intriguing. At elevated $T$, electron-electron (e-e) collisions can provide an additional contribution $1/f\tau_\mathrm{ee}\equiv\gamma T^2$ to the damping parameter $\alpha$ in Eq.~(\ref{Vph}) \cite{hatke:2009a,ryzhii:2004,mamani:2008,dmitriev:2009b}.
The relevant lifetime $\tau_\mathrm{ee}$, responsible for the effective broadening of LLs, is given by the Fermi-liquid e-e scattering rate \cite{chaplik:1971,giuliani:1982}, $\tau_\mathrm{ee}^{-1}= c T^2/\varepsilon_\mathrm{F}\hbar$, where $\varepsilon_\mathrm{F}$ denotes the Fermi energy,
and constant $c$ of order unity includes the logarithmic and numerical factors \cite{dmitriev:2005,dmitriev:2009b}. Our data presented in Fig.~\ref{Fig2}d reveals that this effect dominates the $T$-dependence of TIMO in the most part of the studied interval of $T$. Indeed, at $T\sim10$~K the amplitude of TIMO around $B=B_\mathrm{F}$ precisely follows the exponential fit $\exp(-\gamma T^2)$. Moreover, the value of
$\tau_\mathrm{ee}\simeq (57\,\mathrm{K}/T)^2$\,ps  extracted from this fit conforms with both experimental values reported for graphene\cite{kumar:2017} (Supporting Information) and the above theoretical estimate (with a reasonable value of $c\simeq 5.6$) \cite{polini:2014,principi:2016}. The above analysis suggests that the parameter $A$ in Eq.~(\ref{Vph}) remains approximately independent of $T$ in the range $T>10$\,K. Such behavior, consistent with the $\exp(-\gamma T^2)$ decay, is characteristic for the displacement mechanism described above \cite{hatke:2009a,ryzhii:2004} (Supporting Information). We also note, that recently observed magnetophonon oscillations (resonant phonon-assisted inter-LL transitions) in graphene~\cite{kumaravadivel:2019,greenaway:2019} also exhibited similarly slow $T$-decay; the latter can potentially be accounted for by e-e scattering as well \cite{hatke:2009b}.


It is instructive to point out that the relevant Fermi energy $\varepsilon_\mathrm{F}\sim 200$~meV in graphene is $\sim 20$ times larger than the standard values in GaAs-based heterostructures used for MIRO measurements. Together with the $\sim 10$ times larger frequency $f=0.69$ THz, this explains why the decay parameter $\gamma\propto 1/f\varepsilon_\mathrm{F}$ is more than 100 times smaller in graphene.

To conclude, we have demonstrated the emergence of strong magnetooscillations in graphene exposed to THz radiation. The oscillations were found to have a common origin with MIRO phenomena observed in 2DES with parabolic spectrum yet they emerge at much higher $f$, persist above liquid nitrogen temperatures and their fundamental frequency is tunable by the gate voltage. The anomalously slow $T$-decay of the observed oscillations compared to other 2DES was demonstrated to be due to a slower rate of e-e scattering responsible for the broadening of LLs.  As an outlook, we note that the linear growth of the oscillation amplitude with increasing power can offer an intriguing opportunity to explore further radiation-driven effects. In particular, the observation of zero resistance states~\cite{mani:2002,zudov:2003,Dorozhkin:2011} in THz-driven graphene together with nonlinear response of Dirac fermions~\cite{raichev:2020} may pave the way for a deeper understanding of the rich spectra of nonequilibrium phenomena in 2DES. Furthermore, due to the resonant character and electrical tunability of the observed photoresponse, our devices can be envisioned as a building block for novel optoelectronic devices.

%

\section{Acknowledgments}

The support from the Deutsche Forschungsgemeinschaft (DFG, German Research Foundation) - Project 
GA501/14-1,  the Volkswagen Stiftung Program (97738), the IRAP program of the Foundation for Polish Science (grant MAB/2018/9, project CENTERA) is gratefully acknowledged. The research was also partially supported through the TEAM project POIR.04.04.00-00-3D76/16 (TEAM/2016-3/25) of the Foundation for Polish Science.
Work at MIT was partly supported through AFOSR grant FA9550-16-1-0382, through the NSF QII-TAQS program (grant number \#1936263), and the Gordon and Betty Moore Foundation EPiQS Initiative through Grant GBMF4541 to PJH. This work made use of the Materials Research Science and Engineering Center Shared Experimental Facilities supported by the National Science Foundation (NSF) (Grant No. DMR-0819762). D.A.B. acknowledges support from MIT Pappalardo Fellowship. The authors thank valuable  discussions with D. Svintsov and L. Levitov. 


\section{Notes}
The authors declare no competing financial interest. E.M., D.A.B. and I.A.D. contributed equally to this work.

\bibliography{bibfileTIRO}
.
\newpage

\begin{widetext}
\setcounter{figure}{0}
\renewcommand{\thesection}{}
\renewcommand{\thesubsection}{S\arabic{subsection}}
\renewcommand{\theequation} {S\arabic{equation}}
\renewcommand{\thefigure} {S\arabic{figure}}
\renewcommand{\thetable} {S\arabic{table}}

\section{Supplementary Information}
\subsection{Device fabrication}
Our encapsulated devices were made following a hot-release method introduced in~\cite{Hot-transfer_NComm}. We first mechanically exfoliated 
monolayer graphene, graphite (10 nm thick) and hBN crystals ($<50$ nm thick). The selected crystals were transferred on top of each other using a Polycarbonate/Polydimethylsiloxane stamp attached to a micromanipulator to obtain a graphite-hBN-graphene-hBN heterostructure. The resulting stack was  released on top of an oxidized silicon wafer above the glass transition temperature of the PC membrane (180$^\circ$ C). After this, the 
heterostructure was patterned using standard electron beam lithography to first define contact regions. Reactive ion etching (RIE) was then applied to selectively remove top hBN layer unprotected by the lithographic resist, leaving trenches for electrical contacts. We deposited 3 nm of Cr and 70 nm of gold using thermal evaporation in high vacuum. Next, we used the same lithography and etching techniques to define the device Hall-bar geometry (Fig. 1a of the main text).

\subsection{Frequency dependence of THz-induced magnetooscillations}
Figure~\ref{FigSf-dep} compares the $V_\mathrm{ph} (B)$ dependencies measured in one of our devices in response to $f=0.69$ and $f=1.63$ THz radiation. Pronounced $1/B$-periodic TIMO were observed for both frequencies although the signal amplitude was significantly larger at lower $f$. 

\begin{figure}[!h]
	\centering
	\includegraphics[width=0.35\textwidth]{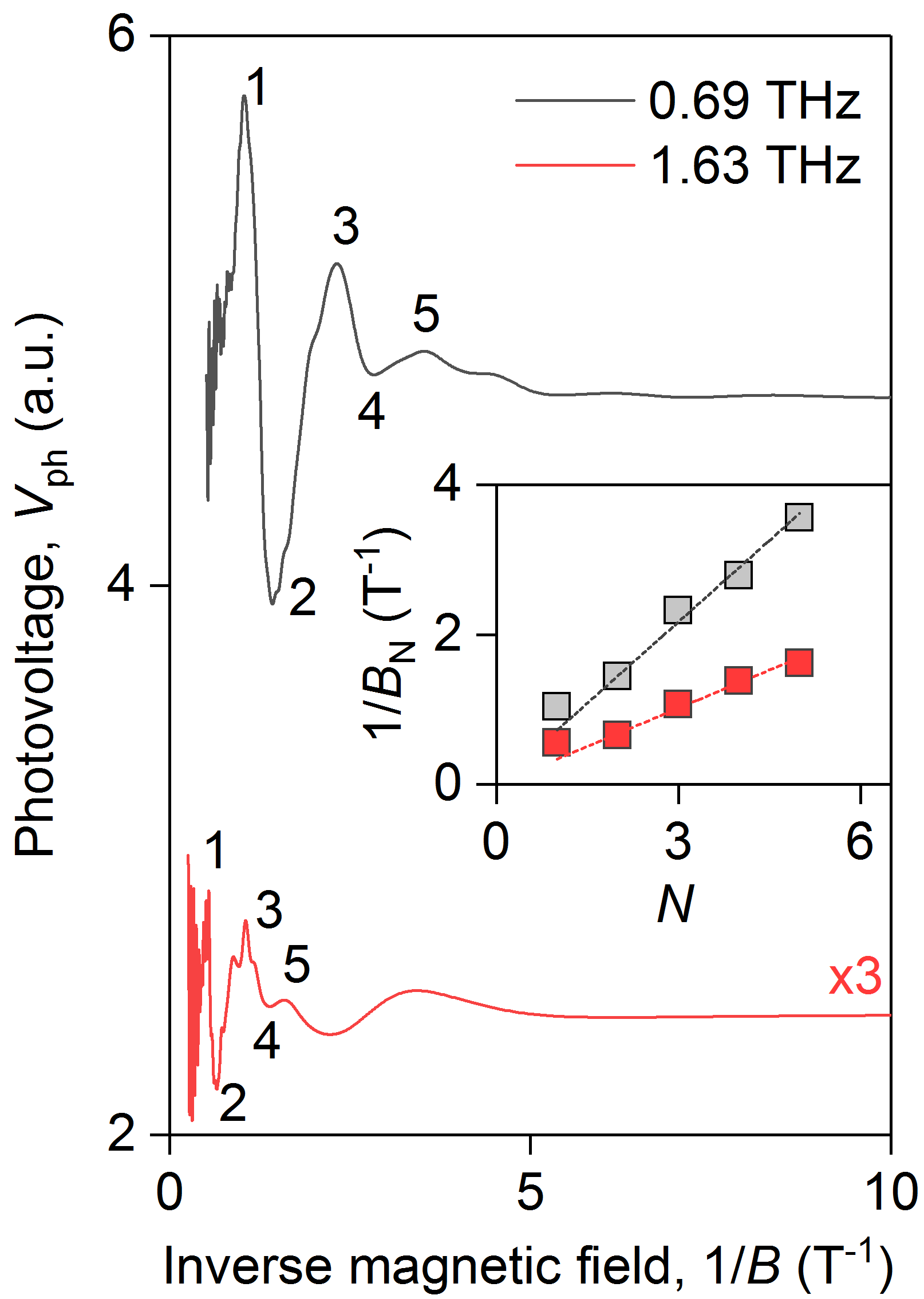}
	\caption{Examples of $V_\mathrm{ph}$ as a function of inverse magnetic field  in response to $f=0.69$ THz and $f=1.63$ THz radiation at fixed carrier density $n=2.75\times 10^{12}$ cm$^{-2}$. Note, the amplitude of the THz-driven $V_\mathrm{ph}$ oscillations recorded at $f=1.63$ THz is significantly lower than that obtained at $f=0.69$ THz. Inset: Inverse magnetic field corresponding to the extrema with number $N$. Dashed lines: linear fit yielding $B_\mathrm{F}=\frac{1}{2}[\frac{d (1/B_\mathrm{N})}{d N}]^{-1}$ reported in Fig.~2c of the main text. $T=4.2$~K.}
	\label{FigSf-dep}
\end{figure}

\subsection{Temperature dependence of THz-induced magnetooscillations}
Figure~\ref{FigSTdep} shows the $V_\mathrm{ph} (B)$ dependencies measured in one of our devices at varying $T$. As one can see from the inset to Fig.~\ref{FigSTdep}, TIMO remain visible up to 90 $K$. The amplitude of the first oscillations period was obtained as a half difference between the respective maxima and minima. In Fig.~\ref{FigSTdep}b we replot the data from Fig.~\ref{Fig2}d of the main text in log-lin scale and show that it accurately follows the $\mathrm{exp}(-\gamma T^{2})$ over the entire $T$-range above 10~K. Using the determined values of $\gamma$, we also plot the corresponding scattering time for electron-electron collisions, $\tau_\mathrm{ee}=1/\gamma f T^2$, which shows remarkable agreement with previous experiments \cite{kumar:2017,polini:2014,principi:2016}. This indicates, that the dominant source of TIMO damping is e-e scattering-induced broadening of LLs.  

\begin{figure}[!h]
	\centering
	\includegraphics[width=0.75\textwidth]{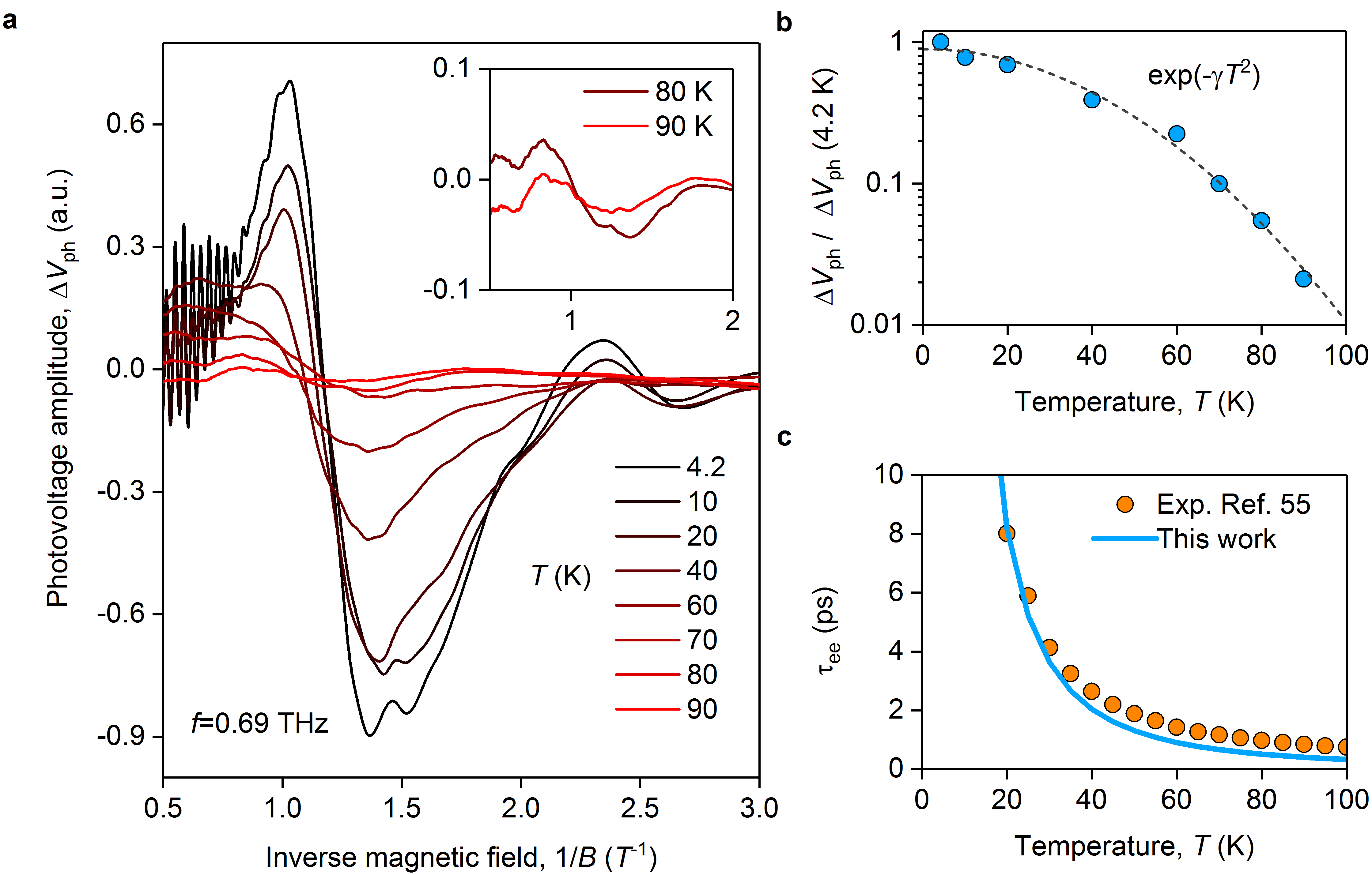}
	\caption{(a) Temperature dependence of $V_\mathrm{ph}$ as a function of inverse magnetic field recorded in response to $f=0.69$ at fixed carrier density $n=2.75\times 10^{12}$ cm$^{-2}$. Inset: Zoomed-in $V_\mathrm{ph}(1/B)$ dependencies at $T=80$ and $90$~K. (b) Normalized to $4.2$~K amplitude of TIMO $\Delta V_\mathrm{ph} / \Delta V_\mathrm{ph}(4.2~\mathrm{K})$ as a function of $T$ plotted in log-lin scale. Dashed line: Fit to $\mathrm{exp}(-\gamma T^2)$ decay with $\gamma\approx4.4\times10^{-4}$ K$^{-2}$. (c) Blue line: Electron-electron scattering time, $\tau_\mathrm{ee}$, as a function of $T$ obtained from $\tau_\mathrm{ee}=1/\gamma f T^2$ using $\gamma$ from the fit in (b). Symbols: Results of previous measurements of $\tau_\mathrm{ee}$ from Ref.~\cite{kumar:2017}.  }
	\label{FigSTdep}
\end{figure}

The analysis of the $T$-decay of TIMO, presented in the main text, suggests that the parameter $A$ in Eq.~(\ref{Vph}) remains approximately independent of $T$ in the range $T>10$\,K. Such behavior, consistent with the $\exp(-\gamma T^2)$ decay, is characteristic for the displacement mechanism of MIRO \cite{hatke:2009a,ryzhii:2004}, which is expected to dominate at high $T$ corresponding to $\tau_\mathrm{tr}\gg\tau_\mathrm{ee}$ \cite{dmitriev:2009b}. In the opposite limit $\tau_\mathrm{tr}\gg\tau_\mathrm{ee}$, the photoresponse is expected to be governed by the inelastic mechanism \cite{dmitriev:2005}, with $A\propto\tau_\mathrm{ee}\propto T^{-2}$ leading to a stronger decay $T^{-2}\exp(-\gamma T^2)$ \cite{dmitriev:2009b, studenikin:2005, studenikin:2007, wiedmann:2010a}. With the above rough estimates for the involved time scales, the transition between the two regimes is expected to happen at $T\sim 30$\,K, while the actual data only indicates a weak deviation from the $\exp(-\gamma T^2)$ decay at the lowest $T$. This apparent discrepancy may reflect the specifics of the thermalization processes in the graphene structures and
warrants further focused studies at low temperatures.

\end{widetext}
\end{document}